\documentstyle[12pt,epsf]{ioplppt}          % use this for preprint style
\begin{document}
\phantom{u}\hfill BI-TP/97-21\\
\phantom{u}\hfill nucl-th/9707020\\
\jl{4}
\title{Chemical equilibration of strangeness}

\author{Josef Sollfrank\ftnote{1}{email:
sollfran@physik.uni-bielefeld.de}}

\address{Department of Physics, University of Bielefeld, 
Universit\"atsstr., D-33615 Bielefeld, GERMANY} 

\begin{abstract}
Thermal models are very useful in the understanding of particle
production in general and especially in the case of strangeness.
We summarize the assumptions which go into a thermal model calculation 
and which differ in the application of various groups. We compare
the different results to each other. Using our own calculation we 
discuss the validity of the thermal model and the 
amount of strangeness equilibration at CERN-SPS energies. Finally
the implications of the thermal analysis on the reaction
dynamics are discussed.
\end{abstract}

%
%  Uncomment out if preprint format required
%

\vspace*{3cm}\noindent
\begin{minipage}{15cm}
Invited talk given at the {\it Int. Symposium on Strangeness in Quark 
Matter 1997}, Santorini (Greece), April 14 -- 18, 1997
\end{minipage}

\maketitle

\section{Introduction}
Strange particles and other heavy flavors play always a special 
role in the analysis of hadronic collisions since they carry a new 
quantum number not present in the incoming nucleons or nuclei. 
Therefore they are considered as one of the most promising tools for
learning more about the dynamics of heavy ion collisions. Especially,
their use as a signature for Quark-Gluon Plasma (QGP) formation was
proposed long ago \cite{Rafelski81,Rafelski82}. The argument is
based on the different time scale for equilibration of strangeness
due to the reduced kinematic threshold in the QGP \cite{Koch86}. 
There is a clear enhancement of strange particle production in 
heavy ion collisions compared to nucleon-nucleon collisions 
\cite{Kinson95,Gazdzicki95a}. Especially, the multistrange baryon
yields increase drastically from $p+p$ to $A$+$A$ \cite{diBari95}.

This interpretation of the measured abundances in favor of
a QGP formation has basically two strategies. First, dynamical
arguments show that the lifetime of a QGP is enough to
reproduce the (still not complete) strangeness equilibration
\cite{Letessier94a,Rafelski96}. Second, the strange particle
ratios are compatible with a sudden disintegrating QGP
\cite{Letessier94b,Letessier95}. However, the interpretation 
is still controversial \cite{Sollfrank95} and alternative explanations 
without QGP formation are also successful 
\cite{Sorge94,ToporPop95,Capella96}.

The various microscopic production models for strange particles
in heavy ion collisions go from
perturbative calculations on parton level \cite{Rafelski96}
to string-string interactions (color ropes) \cite{Sorge94} to 
hadronic rescattering mechanisms \cite{Capella96}. We leave out
the complicated question of the dynamics of strange quark production
and address the simpler question of whether strangeness production
already saturates in heavy ion collisions, i.e. it reaches chemical
equilibrium in the final state. This investigation concentrates
only on the final freeze-out and it implicitly assumes that the particle
production may be described in a statistical or even thermal model. 
Therefore one has first to address the more general question whether 
the overall particle production is given by a thermal production 
mechanism. The answers will in general depend on the particle species, 
the center of mass energy and the volume.

The recent success in the statistical interpretation of particle production 
in elementary reactions like $e^+ + e^-$ \cite{Becattini96a}  and $p + p$ 
($\bar{p} + p$) \cite{Becattini97a} as well as in heavy ion reactions 
at various 
energies \cite{BraunMunzinger95,BraunMunzinger96} triggered a revival of the 
thermal model applications. Therefore we will concentrate to review the 
status of the thermal model for particle production in general and in the 
special case of strangeness. In section 2 we will present a 
hitch hikers guide through thermal models addressing various
points which differ in the application of the model by various groups.
Section 3 is devoted to the discussion of our own calculation 
done for CERN-SPS energies. In section 4 we discuss the implications 
from the thermal analysis of particle production and give a general
conclusion.

\section{The thermal model}

A lot of publications have addressed the question of particle
production within the thermal model. This can be seen in table 
\ref{summary} where only recent ones have been included.
The list is certainly not complete due to ignorance or due to
series of publications of one group where we took only the most recent 
one. Even if the
thermal formalism is used for the same collision system the results
vary (see table \ref{summary}) because the model is applied in
various approximations and extensions which we would like to discuss
first. This will help to understand the differences in the calculations
and the conclusions about the physics of these reactions.

\begin{table}
\caption{Summary of calculations addressing the chemical
freeze-out in various collisions. \label{summary}} 

\begin{indented}
\lineup
\item[]\begin{tabular}{@{}llllll}
\br                              
                       &  $\sqrt{s}$      &               &              & 
                  & \\ 
  collision            & (MeV)            & $T$ (MeV)     &$\mu_B$ (MeV) & 
$\gamma_s$       & ref. \\ 
\mr
$e^+$ + $e^{-}$        & 29.0             & $196\pm7     $ & -           & 
1            & \cite{Hoang88}\\
$e^+$ + $e^{-}$        & 91.5             & $261\pm9     $ & -           & 
1            & \cite{Hoang94}\\
$e^+$ + $e^{-}$        & 29.0             & $163.6\pm3.6 $ & -           & 
$0.724\pm0.045$ & \cite{Becattini96b}\\
$e^+$ + $e^{-}$        & 91.5             & $160.6\pm1.7 $ & -           & 
$0.675\pm0.020$ & \cite{Becattini96b}\\
\hline
p+p             & 19.5             & $161\pm31    $ & $200\pm37$  & 
$0.22\0\pm0.05  $ & \cite{Sollfrank94}\\
p+p              & 19.5             & $190.8\pm27.4$ & -           & 
$0.463\pm0.037$ & \cite{Becattini97a}\\
p+p              & 27.5             & $169.0\pm2.1 $ & -           & 
$0.510\pm0.011$ & \cite{Becattini97a}\\
\hline
Si+Au            & \05.3            & $100.2$      & $559.5$  &
%$89.9$    &  
1              &\cite{Davidson91a}\\
Si+Au            & \05.3            & $127\pm8$      & $485\pm70$  &
$0.5\pm0.2^{\rm a}$ &\cite{Letessier94c}\\
Si+Au            & \05.3            & $140\pm5$      & $555\pm33$  &
$ 1 $         &\cite{Panagiotou96}\\
Si+Au(Pb)        & \05.3            & $130\pm10$     & $540$       &
$ 1 $         &\cite{BraunMunzinger95}\\
Si+Au            & \05.3            & $110\pm5$      & $540\pm20$  &
$ 1 $         &\cite{Cleymans97a}\\
\hline
Au+Au        & \04.7            & $100\pm4$      & -  &
$ 1 $         &\cite{Cleymans97b}\\
\hline
S+S              & 19.5             & $170$     & $257$  &
$1$   &\cite{Davidson91b}\\
S+S              & 19.5             & $197\pm29$     & $267\pm21$  &
$1.00\0\pm0.21$   &\cite{Sollfrank94}\\
S+S              & 19.5             & $185     $     & $301$       &
$1$           &\cite{Tiwari96}\\
S+S              & 19.5             & $192\pm15$      & $222\pm10$ &
$1$           &\cite{Panagiotou96}\\
S+S              & 19.5             & $182\pm9$      & $226\pm13$  &
$0.73\0\pm0.04$   &\cite{Becattini97b}\\
S+S              & 19.5             & $202\pm13$      & $259\pm15$  &
$0.84\0\pm0.07$   &\cite{Sollfrank97b}\\
\hline
S+Ag             & 19.5             & $191\pm17$      & $279\pm33$ &
$1$   &\cite{Panagiotou96}\\
S+Ag             & 19.5             & $180.0\pm3.2$      & $238\pm12$  &
$0.83\0\pm0.07$   &\cite{Becattini97b}\\
S+Ag             & 19.5             & $185\pm8$      & $244\pm14$  &
$0.82\0\pm0.07$   &\cite{Sollfrank97b}\\
\hline
S+Pb       & 19.5             & $172\pm16$     & $292\pm42$ &
$1$           &\cite{Andersen94}\\
S+W       & 19.5             & $190\pm10$     & $240\pm40$ &
$0.7$           &\cite{Redlich94}\\
S+W       & 19.5             & $190$          & $223\pm19$  &
$0.68\0\pm0.06$   &\cite{Letessier95}\\
S+W        & 19.5             & $196\pm9$      & $231\pm18$  &
$1$           &\cite{Panagiotou96}\\
S+Au(W,Pb)       & 19.5             & $165\pm5$      & $175\pm5$  &
$1$           &\cite{BraunMunzinger96}\\
S+Au(W,Pb)       & 19.5             & $160$      & $171$  &
$1$           &\cite{Spieles97}\\
S+Au(W,Pb)       & 19.5             & $160.2\pm3.5$      & $158\pm4$  &
$0.66\pm0.04$           &\cite{Sollfrank97b}\\
\br
\end{tabular}
\\
\item[] $^{\rm a}$ only guessed.
\end{indented}
\end{table}

\subsection{Basic particle yields}
In the thermal model the particle yields are given by a temperature
$T$ and a volume $V$ common for all particles. If one considers only
particle ratios then in most of the applications the ratio is 
independent of $V$ as it was shown in \cite{Cleymans97c}. 
In addition the abundance of a
particle depends on its conserved quantum numbers. This
is either regulated by chemical potentials in the grand canonical
description or by restricting the partition function only to states
which have the same quantum number as the fireball, i.e.~canonical treatment.
The basic expression for the abundance $N_j$ of a particle of species $j$ 
is given by
\begin{equation}
N_j = \lambda_j \frac{\partial \ln Z(T,V,...)}{\partial \lambda_j}.
\end{equation}
All models fulfill this basic requirement with the exception of
the work of Hoang for $e^+$ + $e^-$ collisions \cite{Hoang88,Hoang94}.
Some empirical formula for the particle yields is used,
which is badly justified. Therefore all results in \cite{Hoang88,Hoang94}
should be taken with care.

\subsection{Statistic}
It is obvious that one should use quantum statistics for 
calculating the partition function, i.e. to use 
Bose and Fermi statistics. However, for practical
reasons one usually switches to the Boltzmann approximation.
The error for pions is at $T = 150$ MeV ($T = 200$ MeV) 9.4\%
(11.3\%), respectively and for kaons at the same temperature
0.5\% (1.1\%), respectively. This estimate suggest to use Bose
statistic for pions while for all other heavier particles the Boltzmann
approximation is valid. Note, that for entropy and pressure
the differences between Boltzmann statistic and Bose/Fermi statistic
are larger. Going to very low temperatures and/or high densities
the use of the right statistic is unavoidable \cite{Lee93}.  

\subsection{Canonical vs grand canonical}
The strong interaction conserves exactly the quantum numbers charge $Q$,
baryon number $B$ and strangeness S. This has to be taken care
in a statistical approach and therefore the question arises
which statistical ensemble concerning these quantum numbers one has to use
\cite{canonical,Redlich80,Cleymans97b}. As a first estimate one usually
considers the fluctuations $\Delta O $ of a conserved quantity $O$ 
in the grand canonical treatment \cite{Pathria72}
\begin{equation}\label{fluctuation}
\Delta O = \sqrt{\langle O^2 \rangle - \langle O \rangle^2} 
\propto \sqrt{N},
\end{equation}
where $N$ is the number of all particles carrying a non zero quantum number
$O$. In order to be better than 10\% this suggest to use canonical
treatment when the number of corresponding particles is below 100.
However, in practice -- depending on the observable calculated --
already a much smaller total amount $N$ of particles is enough to
have accuracy better then 10\% using the grand canonical ansatz. This 
was shown in \cite{canonical} and also recently in \cite{Cleymans97b}, 
where particle ratios already seem to saturate for number of participating 
baryons of around 30--40.

In the applications for heavy ion collisions the grand canonical
treatment is justified when single particle yields are addressed
\cite{Cleymans91}. We tested the difference in the resulting
thermal parameters using canonical and grand canonical treatment 
for strangeness. We saw only minor differences between both calculations
for ultra-relativistic heavy ion collisions.
However, for strange particle correlations the use of the canonical 
ansatz is recommended \cite{Cleymans91}.

If the thermal model is applied to inelastic two body collisions
at high energies the analysis should be done using the canonical ensemble. 
Among the calculations in table \ref{summary} this was only performed by
Becattini \cite{Becattini96b,Becattini97a}. All other calculations
concerning $p$ + $p$ and $e^+$ + $e^-$ have to be 
taken with care. For example, we made a least mean square fit ($\chi^2$-fit)
using the grand canonical ensemble for charge, baryon number and
strangeness for particle production in p+p collisions at 
$\sqrt{s} = 27.5$ GeV. The experimental input data are the same
as in the work of Becattini \etal \cite{Becattini97a} and the resonances
included are very much the same. The fit to the data gets worse
but the fit temperature stays very much the same and turned
out to be $T = 160$ MeV but the strangeness suppression 
$\gamma_s$ (see Section \ref{offequilibrium}) is reduced
to $\gamma_s = 0.35$ compared to $\gamma_s = 0.51$ in 
\cite{Becattini97a}. The canonical treatment suppresses the
strangeness production due to associate pair production leading
naturally to lower strangeness yields as compared to elementary reactions.

The chemical potentials in the grand canonical approach are first
of all Lagrangian multipliers for the conserved charge. However, the
strange quark chemical potential $\mu_{\rm s}$ plays a 
special role in the interpretation of strange particle abundances
\cite{Rafelski91}. It is zero in a strangeness neutral QGP
but has usually non-zero values in an equilibrated hadron gas. 
The $\mu_{\rm s} = 0$ line in the $T$-$\mu_{\rm B}$ plane calculated
for an equilibrated hadron gas is close to the expected QGP phase 
transition line \cite{Asprouli95,Panagiotou96}. 
This similarity leads to speculations about the
meaning of the $\mu_{\rm s} = 0$ line which were extensively discussed
in \cite{Asprouli95,Panagiotou96}.

\subsection{Isospin}
Isospin breaking effects become important when target and projectile 
nuclei are large. In most of the thermal models isospin is
neglected. We may estimate the effect of isospin breaking to be of
order $1 - 2Z/A$. This leads in S+Au collisions to a correction
of order 10 \% and for Pb+Pb of order 20\%. Therefore one should
take charge and baryon number separately into account especially 
for isospin sensitive quantities
like $\pi^+/\pi^-$ or $K^+/K^-$. All above listed models in table
\ref{summary} neglect isospin in the case of a heavy target.
As an example one may discuss the $\pi^+/\pi^-$ ratio which is
different from one in the low $p_T$-region of Au + Au \cite{Ahle95}
and Pb + Pb \cite{Boggild96} collisions by chemical equilibration
arguments. This ratio is sensitive to isospin violating weak
decays as discussed in \cite{Arbex97}. But in addition the isospin
asymmetry in the colliding nuclei plays an important role and this
was neglected in \cite{Arbex97}.

The isospin SU(2) may be treated exactly like in \cite{Redlich80} or to
a very good approximation as an U(1) $\times$ U(1). Then one usually
conserves baryon number and charge or on valence quark level
the net number of  u-quarks and net number of d-quarks.
In a grand canonical treatment it is equivalent to introduce chemical
potential for baryon number and charge or for up and down quarks.

\subsection{Finite volume correction}
Calculating the partition function one switches from the summation 
over the states to the integral representation
\begin{equation}
\sum\limits_{k}^{\rm states} \longrightarrow
\int \d^3 \vec{k} \; g(\vec{k}) = 4\pi \int \d k \; k^2 g(k),
\end{equation}
where $g(k)$ is the density of state. If the volume is small $g(k)$ is 
very different from the infinite volume limit $V/(2\pi)^3$ and often
approximated by \cite{Hill53,Pathria72}
\begin{equation}\label{gk}
g(k) = \frac{V}{(2\pi)^3} - \frac{S}{32 \pi^2 k} + \frac{L}{32 \pi^2 k^2},
\end{equation}
where $V$ is the volume $S$ the surface and $L$ the circumference
of the box. This is only an approximate formula derived for
Dirichlet boundary conditions and the error is of the order of
the last term. This kind of correction was applied in 
\cite{BraunMunzinger95,BraunMunzinger96,Davidson91a,Davidson91b}.

Going to $p$ + $p$ collisions one might think that this kind
of correction might be very important. However, we like to argue that
in this case and also for ultra-relativistic heavy ion collisions
it is reasonable to use the continuous density of state for the
following reason. In equation (\ref{gk}) it is assumed that the states 
have to fulfill Dirichlet boundary conditions on the edge of
the interaction volume. This is only true for 
an infinite high potential well. In reality there is
no such large binding force for the particles squeezing its wave function to
the reaction volume. The particle energies are much higher than a
nuclear binding potential. Therefore we think that the use of the
continuous density of state is appropriate as long as the thermal
energies are above a mean field potential which serves as a confining box.

\subsection{Resonances}
It is now commonly agreed that the particle production is dominated
by resonance production mechanisms. Therefore the resonance states 
are included in the partition function of all calculations
in table \ref{summary} except for \cite{Hoang88,Hoang94}. On the
other hand their decay is sometimes neglected 
\cite{Panagiotou96,Tiwari96} when calculations are
compared to experimental data. Some of the publications restrict their 
discussion to strange baryon ratios. Then the resonance 
contribution may be restricted to the inclusion of the $\Sigma^0$ decay
which feeds into the $\Lambda$ yield. This gives a reasonable estimate 
for the fugacities. If, however, kaons are included in the chemical analysis
one has to consider the sizeable feeding from resonances and its 
omission is hardly comprehensible. In the treatment of resonances 
the following items are important.

\subsubsection{Number of resonance states}
The modeling of a steady state hadron gas needs the input of all
resonance states leading to the bootstrap description of 
thermodynamics of strong interacting matter \cite{Hagedorn71}.
As a result one gets an exponential increasing density of
resonance states 
\begin{equation}\label{bootstrap}
\rho(m) \propto m^{a}\exp(m/T_{\rm H})
\end{equation}
with the
consequence of a limiting temperature $T_{\rm H}$, called Hagedorn
temperature. In the applications for hadronic collisions it is assumed that
only a finite range in the resonance mass spectrum is equilibrated.
The finite volume of the fireball leads to a finite total energy giving
an upper limit in the mass spectrum to be considered. This estimate 
corresponds practically to an infinite mass spectrum. It arises the
question whether already a much lower cut in the resonance spectrum
can be justified. If chemical equilibrium is build up via secondary
interactions than the limited life time of the fireball leads to
a limited amount of equilibrated resonance states. If, however,
the hadronization in elementary p + p collisions follows the 
chemical equilibrium abundances -- and this seemed to be the case 
\cite{Becattini97a} -- then it is hard to 
argue for the omission of the higher mass states, especially if the 
temperature is around 200 MeV. Nevertheless the known resonance
states fade at masses around 2 GeV and therefore the calculations
have to be restricted to a finite number of states.

The applied cut in the resonance states of the various
groups is arbitrary and given by practical considerations. It
has been shown that there is a small influence on the extracted
thermal parameters on the cut in the resonance spectrum 
\cite{Sollfrank94,Becattini96a,Becattini97a}. The temperature fitted
to measured particle ratios shows a maximum at a resonance 
cut-off at $\approx 1.5$ GeV.
It is interesting to note that fits to the experimental known
density of states $\rho(m)$ by the bootstrap formula in
equation (\ref{bootstrap})
start to deviate at the same mass of 1.5 GeV \cite{Tounsi94}.
Therefore one should be aware that thermal fits at temperatures
higher than $\approx 170$ MeV are biased from the resonance spectrum
which is taken into account.

\subsubsection{Branching of resonances}
Not only the poor knowledge of resonance states around 2 GeV has some
influence on the analysis for high temperatures, but also the branching
ratios. Already at 1.5 GeV they are starting of getting basically unknown
and the various groups use some ``educated guess'' \cite{Cleymans97a}.

\subsubsection{Resonance width}
Usually the width of a resonance is neglected. This means that in 
the Boltzmann factor $\exp(-\sqrt{m^2 + p^2}/T)$ the mean mass $\bar{m}$
of a resonances is taken.
This assumption was made in nearly all calculations of table \ref{summary}.
For broad resonances this is a bad approximation.
One suggestion to improve this is to distribute the mass states
of a resonance according to a Breit-Wigner form
\cite{Becattini96a,Becattini97a,Becattini97b,Sollfrank97b}.
This means that
the mass shell constraint in the Lorentz invariant momentum integration
for the partition function is replaced by \cite{Sollfrank91}
\begin{eqnarray}\label{width}
\int \d^4 p \; \delta\left( p_\mu p^\mu - m_0^2\right) 
\theta \left(\sqrt{p_\mu p^\mu}\right) 
&& \nonumber \\ 
\longrightarrow 
\int \d^4 p \; \frac{m_0\Gamma}{\left(p_\mu p^\mu - m_0^2\right)^2 
- m_0^2\Gamma^2} \;
\theta\left(\sqrt{p_\mu p^\mu} - m_{\rm thres}\right), &&
\end{eqnarray}
where $\Gamma$ is the width of the resonance and 
$m_{\rm thres}$ a threshold for the resonance production.
The inclusion of the width is important for the yield of the 
$\rho$-meson. It turned out to improve the fits
in p + p collisions \cite{Becattini97a}.

\subsection{Repulsive interaction}
The experimental particle yields and ratios are determined at (chemical)
freeze-out. One usually assumes that the system is already such diluted 
that the interactions have effectively cease to exist. Then the problem
of residual interactions don't occur. However, in the analysis of SPS 
energies the chemical freeze-out temperature seems to be around 160--200 MeV
(see table \ref{summary}). At this temperature the particle density is
still very high and one has to ask how density corrections influence 
the particle yields. In the bootstrap model of Hagedorn the dominant 
part of the attractive strong interaction is effectively taken into account 
by including the higher resonance states \cite{Hagedorn71}. However, 
at high densities the repulsive part have to be accounted for, too. 
It is popular to use an excluded volume correction 
\cite{Hagedorn80,Cleymans86,Rischke91,Uddin94} where the hadrons are 
treated as finite size hard core particles. In the early 
suggestions \cite{Hagedorn80,Cleymans86} the real physical particle density
$n^{\rm phy}$ is related to the ideal or point particle density 
$n^{\rm point}$ by 
\begin{equation}\label{excludedv}
n^{\rm phy} = \alpha^{-1} n^{\rm point},
\end{equation}
where $\alpha$ is either given by the total point particle 
energy density $\varepsilon_0$ and the bag constant $B$, i.e. 
$\alpha = 1 + \varepsilon_0/(4B)$ \cite{Hagedorn80}, 
or by $\alpha = 1 + \sum_j V^0_j n_j^0$ where $V^0_j$ is a 
hard core volume and $n_j^0$ the point particle density 
of species $j$ \cite{Cleymans86}.    
The important point is that such a treatment don't influence 
particle ratios since the factor is common for all particle
species and therefore don't change the thermal analysis. The
volume $V$ which appears as a common factor has to be regarded as
the point particle volume and the physical volume is then 
given by $\alpha V$.

The above described correction is thermodynamically not 
consistent \cite{Hagedorn80,Cleymans86} and improvements have been suggested 
\cite{Rischke91,Uddin94,Kapusta83} (for an extended discussion see 
\cite{Venugopalan92}).
One may divide them basically into mean field approaches like
\cite{Kapusta83} or thermodynamically consistent excluded volume corrections 
like \cite{Rischke91,Uddin94} or both \cite{Rischke91}. 
These models contain additional parameters
which characterize either the hard core size $V^0_j$ or the
mean field coupling $K_j$ of a particle $j$. If $V^0_j$ or $K_j$
are different for various $j$ then they influence the particle ratios.
The only publications in table \ref{summary} which have such an influence 
on particle ratios due to repulsions are 
\cite{Davidson91a,Davidson91b,Tiwari96}. 
In \cite{Davidson91a,Davidson91b} the effect is minimal since the mean 
field coupling $K_B = 680$ MeV fm$^3$ of baryons and anti-baryons is 
similar to the one of mesons $K_M = 600$ MeV fm$^3$. The studies in 
\cite{Tiwari96} show the repulsion effect for the heavy baryons.
The extracted $\mu_B$ in \cite{Tiwari96} differs from 
the other results of table \ref{summary}. Tiwari \etal 
take for the hard core size of a particle the MIT bag model result
of $V^0_j = m_j/(4B)$, i.e. the hard core volume scales with its mass.  
Therefore the massive baryons are additionally suppressed by their
size. This has to be compensated by a higher $\mu_B$. 

\subsection{Off-equilibrium phenomenology\label{offequilibrium}}
Since the life time of a fireball created in heavy ion collisions
is very short and the dynamics is very rapid it cannot be expected that 
the production of particles of all kind follow the equilibrium statistics.
In order to study the deviations from equilibrium quantitatively one 
introduces over-saturation/suppression factors $\gamma$ which measure the
deviations from full equilibrium. For strangeness they were first 
introduced by Rafelski \cite{Rafelski91} in a phenomenological way.
They were defined by
\begin{equation}
\gamma = \frac{\rm actual\;density}{\rm equilibrium\;density}.
\end{equation}
It has been shown \cite{Slotta95} that the thermodynamically correct 
way of defining such a parameter is to define them as fugacity 
\begin{equation}\label{gammas}
\gamma = \exp(\mu/T),
\end{equation}
as it is done in a grand canonical approach. 
This is in accordance with the model
of relative chemical equilibrium. This means that only a subset
of particles is in chemical equilibrium among each other and
the deviation to another set of particles is parametrized by $\gamma$. 
Examples for such applications are the strangeness suppression $\gamma_s$ 
\cite{Rafelski91}, the pion chemical potential 
\cite{Kataja90} or general meson and baryon suppression factors
\cite{Letessier94d}.

With the help of suppression factors one is able to study the
approach to chemical equilibrium.
Some of the calculations in table \ref{summary} allow for such a suppression
and some don't. Since we know from $p$ + $p$ collisions \cite{Becattini97a}
that strangeness is suppressed by roughly a factor of 2 one should
always allow the suppression possibility in heavy ion reactions.
In table \ref{summary} all calculation which don't allow strangeness 
suppression have 1 in the column for $\gamma_s$.

\subsection{Restricted acceptance}

\begin{figure} 
\caption{Pion (solid line) and proton (dashed line) rapidity 
distribution of a static fireball at
a temperature of $T= 150$ MeV normalized to one at central rapidity.
\label{rapidity}}
\hspace*{2.5cm} \epsfxsize 10cm \epsfbox{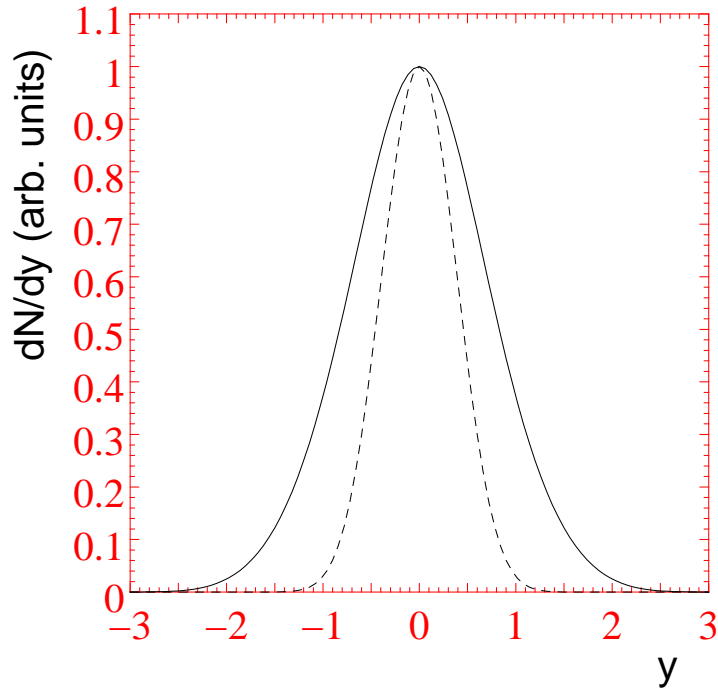}
\end{figure}

If thermal equilibrium is established it is either globally or
locally. In most of the cases as for example in S+S collisions at 
CERN-SPS the rapidity distributions of baryons and pions 
are very different in shape \cite{Bachler94}. This suggest that the 
freeze-out parameters vary locally. One way to demonstrate this 
was assuming a rapidity dependence of the baryon chemical potential 
as done in \cite{Slotta95}. However, the proper treatment of 
locally varying thermal parameters is via hydrodynamics 
\cite{Sollfrank97a}. On the other hand, 
if the differences are small, the use of one global fireball
in the analysis of $4\pi$ data is appropriate and the result should
be understood as a global average.

The most reliable way to analyze particle yields and ratios is 
using $4\pi$ integrated data. This avoids problems
due to kinematic cuts. Addressing particle ratios in restricted
kinematic regions needs a model for the particle spectra, too.
Particle spectra are much more sensitive to the dynamics and
one needs more assumptions and knowledge about the space-time evolution
in longitudinal and transverse direction. Therefore we recommend the
use of $4\pi$ data for the study of chemical equilibration.

Most of the analysis in table \ref{summary} is applied to
particle ratios in a restricted kinematic region. The reason is that 
most of the particle ratios are only measured in a kinematic window. 
This requires 
that the calculation is cut to the experimental acceptance and
the knowledge of the particle spectra is unavoidable. 

We like to demonstrate how much results in principle may depend on 
assumptions about
the longitudinal dynamics at freeze-out. The two extreme scenarios
are a static fireball and the Bjorken boost-invariant scenario
\cite{Bjorken83}. In a static fireball the rapidity distribution using
Boltzmann approximation is given by
\begin{eqnarray} \label{static}
\frac{\d N^{\rm static}}{\d y} (y) &=&  
\frac{V g m^2 T}{2\pi^2}\exp\left[-(m\cosh(y) -\mu)/T\right] \nonumber \\ 
&&\times \left[ \frac{1}{2} + \frac{T}{m\cosh(y)} + 
\left(\frac{T}{m\cosh(y)}\right)^2 \right],
\end{eqnarray}
while in the Bjorken case it is
\begin{equation}\label{bjo}
\frac{\d N^{\rm Bjo}}{\d y} (y) \propto 
\frac{g m^2 T}{2\pi^2} \; \exp(\mu/T) \; {\rm K}_2(m/T).
\end{equation}
It is $V$ the volume $g$ the spin degeneracy and $m$ the mass.
As an example we show in figure \ref{rapidity} the pion and proton rapidity
distributions for a static fireball given by equation (\ref{static}). 
The pion and proton distribution are normalized to one and
therefore its ratio at midrapidity is one. In the Bjorken scenario
the ratio of pions to protons is given by the integration over rapidity
in equation (\ref{static}) resulting in the expression of equation
(\ref{bjo}). In our example it would be
$(dN^{p}/dy)/(dN^{\pi}/dy) = 0.54$. Since the mass of pions and
protons are very different the effect is most pronounced in the
example. We show in Section \ref{results} that in practice the
differences between both scenarios are minor using the example of S+Au
collisions.

The experimental rapidity distributions are broader than given
by a static source (\ref{static}) but not infinite broad like in the
case of Bjorken scaling. Since the experimental width of various 
rapidity spectra is very similar a thermal analysis in
a restricted rapidity range usually assumes Bjorken scaling and
uses equation (\ref{bjo}) for particle yields. 

\section{Results from thermal model analysis \label{results}}

The strength of the thermal model is that most of the particle ratios 
can be explained by only a few parameters. However, from table \ref{summary}
one cannot see how well the thermal model works and where the deviations
start. Therefore we show the results of one calculation in more detail.
We perform a thermal model calculation which has the following 
characteristic \cite{Sollfrank97b}:
\begin{itemize}

\item All hadronic states up to 1.7 GeV in mass are included.

\item Pions follow the Bose statistic while for all other
      hadrons the Boltzmann approximation is used.

\item The resonances are populated including their width according
      to equation (\ref{width}). The Breit-Wigner distribution in mass 
      is restricted to a range of two times the width
      \cite{Becattini96a}.
     
\item When comparing to experimental results we include the feeding
      of resonances and in the case of S+Au we also include the
      $p_T$-cut of the experimental ratios. 

\item For S+S and S+Ag strangeness is treated in the 
      canonical formalism while for S+Au we use the grand canonical
      ensemble. Baryon number is
      regulated via a chemical potential. Isospin symmetry is assumed.

\item No finite size correction and no repulsive interaction is
      included.
\end{itemize}

We analyze experimental particle yields in two ways. On the one hand
we perform a $\chi^2$-fit to $4\pi$ data of S+S and
S+Ag collisions and to central particle ratios in S+Au 
collisions. The second possibility is to display the experimental
particle ratios in the $T$-$\mu_B$ plane and look for overlap regions
of the various bands.

\begin{table}
\caption{Result of a fit to experimental $4\pi$ data
of S+S and S+Ag collisions at CERN-SPS. The data are 
all measured by the NA35 collaboration.
\label{ss}} 

\begin{indented}
\lineup
\item[]\begin{tabular}{@{}lllllll}
\br                              
                       &  S+S      &               &              & 
S+Ag            &          & \\ 
                       &  calculation       &  data        & ref.
                       &  calculation       &  data        & ref.\\
\mr
$h^-$          &   83.7         & $94\pm5$      & \cite{Bachler94}
               &   151          & $160\pm8$     & \cite{Rohrich94}\\
$K^+$          &   12.7         & $12.5\pm0.4$  & \cite{Bachler93}
               &   23.0         &             & \\
$K^-$          &   7.13         & $6.9\pm0.4$   & \cite{Bachler93}
               &   13.3         &             & \\
$K_s^0$        &   9.70         & $10.5\pm1.7$  & \cite{Alber94}
               &   17.8         & $15.5\pm1.5$  & \cite{Alber94}\\
$\Lambda$      &   8.69         & $9.4\pm1.0$   & \cite{Alber94}
               &   14.4         & $15.2\pm1.2$  & \cite{Alber94}\\
$\bar{\Lambda}$&   1.84         & $2.2\pm0.4$   & \cite{Alber94}
               &   2.54         & $2.6\pm0.3$   & \cite{Alber94}\\
$p - \bar{p}$  &   22.6         & $20.2\pm2.0$  & \cite{Bachler94}
               &   38.1         & $34\pm4$      & \cite{Rohrich94}\\
$\bar{p}^{\rm a}$& 1.93         & $1.15\pm0.4$  & \cite{Alber96}
               &   2.99         & $2.0\pm0.8$   & \cite{Alber96}\\
\hline
T (MeV)      &  $202\pm13$        &         & 
             &  $185\pm8$         &         & \\
V (fm$^3$)   &  $81.5\pm39.4$     &         &
             &  $275\pm84 $       &         & \\    
$\gamma_s$   &  $0.84\pm0.07$     &         &
             &  $0.82\pm0.07$     &         & \\ 
$\lambda_q$  &  $1.532\pm0.038 $  &         &
             &  $1.552\pm0.041$   &         & \\
\hline
$\chi^2$/dof &  11.6/4            &         &
             &  6.72/2            &         & \\
\br
\end{tabular}
\\
\item[] $^{\rm a}$ The experimental value is extrapolated to $4\pi$
assuming the same rapidity shape as the $\bar{\Lambda}$.
\end{indented}
\end{table}

In table \ref{ss} we show the result of a $\chi^2$-fit 
to $4\pi$ data from the NA35 collaboration 
(see references in table \ref{ss}). In S+Ag collisions
the assumed isospin symmetry is slightly violated. In order to 
estimate the size of the effect we have 
also done a fit using separate chemical potentials for up and down 
quarks and including the total net charge. In the case of S+Ag
collisions the result is a slightly larger $\lambda_d = 1.572\pm0.053$ 
as compared to $\lambda_u = 1.521\pm0.034$. This is expected by the
larger amount of incoming u-quarks. 

The calculations are very similar to the one of Becattini 
\cite{Becattini97b} but note that in the case of S+S collisions 
a different input set of experimental data is
used. Therefore we get a much higher temperature for S+S while 
for S+Ag both calculations basically agree. The differences between them
are small deviations in the input resonances states,
their branching and the treatment of the $\eta$-$\eta^\prime$ mixing.

Looking at the result in more detail one realizes that the thermal 
fit is not perfect, especially for S+S collisions 
($\chi^2/{\rm dof} = 11.6/4$). The largest
deviations are in the anti-baryon yields. The high absorption cross
section of this particles may explain the deviations.
 
\begin{figure}
\caption{\label{ssfig} 
Experimental particle ratios in the $T$-$\mu_B$ plane for S+S
collisions taking $\gamma_s = 0.84$ and $V = 81.5$ fm$^3$. 
The bands along the abscissa correspond to the following 
ratios (going from left to right):  $K^-/h^-$ (\dotted), 
$\Lambda/(p-\bar{p})$ (\broken),
$\bar{\Lambda}/\Lambda$ (\full), $\bar{p}/(p-\bar{p})$ (\longbroken),
$h^-/(p-\bar{p})$ (\chain) $K^+/K^-$ (\full) and 
$K_s^0/\Lambda$ (\longbroken). The point indicates the result
of the $\chi^2$-fit in table \ref{ss}.}
\hspace*{2.5cm} \epsfxsize 10cm \epsfbox{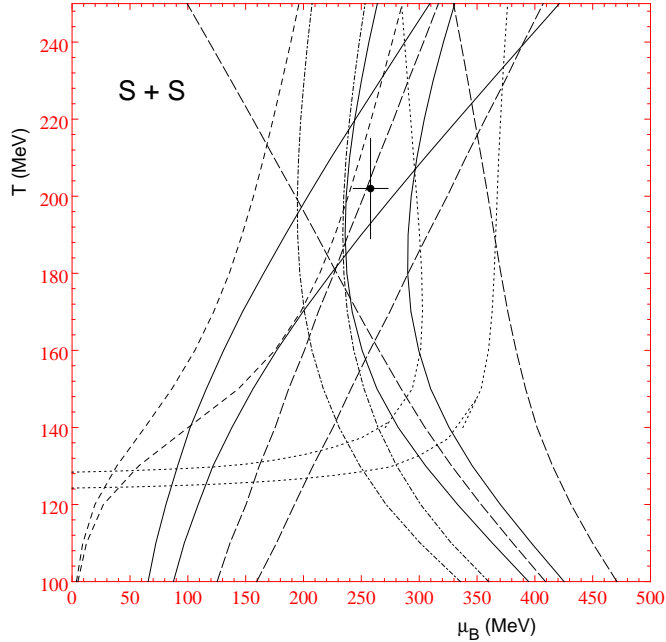}
\end{figure}

A different way of displaying the quality of the thermal model approach 
is shown in figure \ref{ssfig} where in the $T$-$\mu_B$-plane 
various bands indicate experimental particle ratios.
We changed to particle ratios because they are nearly independent of
the volume. Since we use the canonical ensemble for
strangeness we have a small influence of the particle ratios on the
volume. We calculated the used experimental particle ratios from 
table \ref{ss}. The errors of the ratios are determined by 
adding the individual errors quadratically. The bands correspond
to the upper and lower bound on the experimental ratio. 
Note that a fixed volume was used.

In figure \ref{ssfig} we see no real overlap region of all
particle ratios. Especially, the ratios containing the $h^-$ fail to cover
the $\chi^2$-fit point which is given by the filled circle. We like to point
out the possible sign of an enhanced entropy production seen in
the $h^-$ as discussed in \cite{Letessier95,Gazdzicki95b}.
Our present reevaluation of the experimental data confirms this possibility. 
We expect a stronger effect on an enhanced pion production in Pb+Pb collisions.

\begin{figure}
\caption{\label{sagfig}
Experimental particle ratios in the $T$-$\mu_B$ plane for S+Ag
collisions taking $\gamma_s = 0.82$ and $V = 275$ fm$^3$.
The bands along the abscissa correspond to the following 
ratios (going from left to right): $K_s^0/h^-$ (\dotted),
$\Lambda/(p-\bar{p})$ (\broken),
$\bar{\Lambda}/\Lambda$ (\full), $\bar{p}/(p-\bar{p})$ (\longbroken)
and $h^-/(p-\bar{p})$ (\chain). The point indicates the result
of the $\chi^2$-fit in table \ref{ss}. 
}
\hspace*{2.5cm} \epsfxsize 10cm \epsfbox{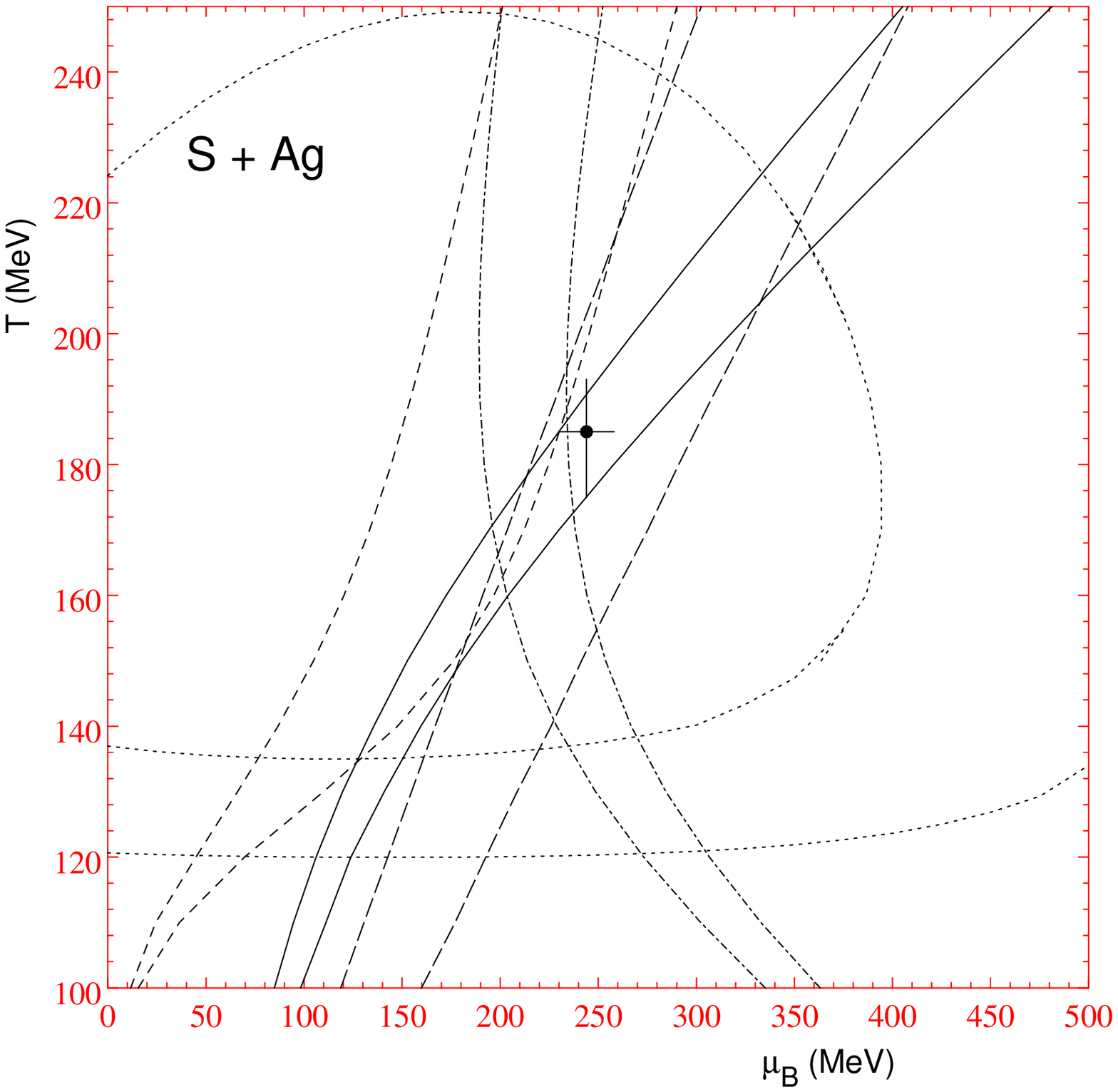}
\end{figure}

In S+Ag collisions the quality of particle production using a thermal
model is similar to the one in S+S as may be seen from the $\chi^2$ 
in table \ref{ss} or in figure \ref{sagfig} where we plotted again
various ratios in the same way as in figure \ref{ssfig}. 
All particle ratios, excluding 
$K_s^0/h^-$, have one common overlap at $T =  170\pm10 $ MeV and 
$\mu_B = 220\pm 10$ MeV. The inclusion of the $K_s^0/h^-$ ratio
in the $\chi^2$-fit moves that result out of the
otherwise common overlap. Excluding $K_s^0$ one gets 
for S+Ag collisions similar freeze-out parameters as for S+Au 
collisions below.
 
\begin{table}
\caption{Result of a fit to experimental central
rapidity particle ratios in S+Au(W,Pb) collisions at CERN-SPS assuming
Bjorken scaling. The extracted thermal parameters are
$T = 160.2\pm3.5$ MeV, $\lambda_q = 1.390\pm0.012$ and 
$\gamma_s = 0.656\pm0.041$. It is $\chi^2/{\rm dof} = 37.7/10$.
\label{sau}} 

\begin{indented}
\lineup
\item[]\begin{tabular}{@{}lllllll}
\br                              
     S+Au                   &  cal. & data& target& rapidity& $p_T$cut& ref.\\
\mr
$D_q^{\rm a}$                     &  0.0782    & $ 0.088\pm0.007$ 
& Pb    & 2.3--3.0  & 0   & EMU05\cite{emu05}\\
$p/\pi^+$                         &  0.188     &   $0.19\pm0.03$ 
& Pb    & 2.6--2.8  & 0   & NA44\cite{Murray94}\\
$\bar{p}/\pi^-$                   & 0.0262      &  $0.024\pm0.009$ 
& Pb    & 2.6--2.8  & 0   & NA44\cite{Murray94}\\
$\bar{p}/p$                       & 0.139       &   $0.12\pm0.02$  
& Pb    & 2.65--2.95& 0 & NA44\cite{Jacak94}\\
$\eta/\pi^0$                      & 0.0816       &  $0.15\pm0.02$ 
& Au    & 2.1--2.9  & 0 & WA80\cite{Albrecht95}\\
$R_K^{\rm b}$                     & 2.03       &  $2.14\pm0.06$
& W     & 2.5--3.0  &1.0& WA85\cite{diBari95}\\
$K^+/K^-$                         & 1.57      & $1.67\pm0.15$
& W     & 2.3--3.0  & 0.9& WA85\cite{diBari95}\\
$K_s^0/\Lambda$                   & 1.21       &  $1.4\pm0.1$
& W     & 2.5--3.0& 1.0 & WA85\cite{Abatzis96}\\
$\bar{\Lambda}/\Lambda$           &  0.203     & $0.196\pm0.011$
& W     & 2.3--3.0& 1.2 & WA85\cite{Evans96}\\
$\Xi^-/\Lambda$                   &  0.0967    & $0.097\pm0.006$
& W     & 2.3--3.0& 1.2 & WA85\cite{Evans96}\\
$\Xi^+/\Xi^-$                     & 0.283     &$0.47\pm0.06$
& W     & 2.3--3.0& 1.2 & WA85\cite{Evans96}\\
$R_\Omega^{\rm c}$                & 0.145     &$0.8\pm0.4$
& W      &2.5--3.0&  1.6&  WA85\cite{diBari95}\\
$\bar{\Omega}/\Omega$             &  0.430    &$0.57\pm0.41$ 
&    W   & 2.5--3.0  & 0 &  WA85\cite{Abatzis93}\\
\br
\end{tabular}
\\
\item[] $^{\rm a}$ $D_q = (h^+ - h^-)/(h^+ + h^-)$ 
\item[] $^{\rm b}$ $R_K = (K^+ + K^-)/K_s^0$
\item[] $^{\rm c}$ $R_\Omega= (\bar{\Omega}+\Omega)/(\Xi^+ + \Xi^-)$

\end{indented}
\end{table}

There are not enough $4\pi$ data on S+Au collisions and 
therefore we switch to particle
ratios. The analysis is inspired by the work of Braun-Munzinger \etal
\cite{BraunMunzinger96}. However we take in our analysis only
a subset of particle ratios from their list, excluding all
ratios which don't cover midrapidity $y_{\rm cm} = 2.65$. 
For the particle yields we use the scaling assumption,
i.e. equation (\ref{bjo}) as it was done in \cite{BraunMunzinger96}. 
In addition we change to the grand canonical ensemble for strangeness.

The result of the $\chi^2$-fit is given in table \ref{sau}.
We reproduce the temperature $T = 160$ MeV as it was assumed in 
\cite{BraunMunzinger96} but we got a slightly lower $\mu_B$ of 158 MeV.
The main difference of our calculation to the one in \cite{BraunMunzinger96}
is that we allow strangeness suppression.
The ratios sensitive to $\gamma_s$ are $\Xi^-/ \Lambda$ and 
$R_\Omega= (\bar{\Omega}+\Omega)/(\Xi^+ + \Xi^-)$.
The resulting low value of $\gamma_s = 0.65$ is not in agreement with
the assumption of full chemical equilibrium for strangeness
as it was assumed in many calculations.

The result of table \ref{sau} shows that a thermal hadron gas model 
even with no complete strangeness equilibration is not able
to reproduce {\it all} experimental data. Some serious deviations
are not in the table like $\bar{\Xi}/\bar{\Lambda} = 0.23 \pm 0.02$
\cite{Evans96} which in the thermal model is given by 
$\bar{\Xi}/\bar{\Lambda} = 0.135$. The disagreement in the multi-strange
baryons might indicate the onset of non-equilibrium physics with
the origin in a QGP formation \cite{Rafelski81}. 
This proposal was recently discussed in depth in \cite{Rafelski96}.

We tested our result against the assumption of Bjorken scaling
in rapidity and did the same fit assuming one static fireball,
i.e. using equation (\ref{static}) at $y=0$. We got a rather similar
result of $T= 158\pm3$ MeV, $\lambda_q = 1.408\pm0.012$ and
$\gamma_s = 0.74 \pm 0.05$. We see in the $\chi^2$-fit no large
dependence on the assumption about the rapidity distribution.

\begin{figure}
\caption{\label{saufig}
Experimental particle ratios in the $T$-$\mu_B$ plane for S+Au(W,Pb)
collisions taking $\gamma_s = 0.66$. In this plot the experimental 
kinematic cuts are not accounted for. The bands are explained along 
the upper part of the figure from left to right: $p/\pi^+$ (\full),
$K_s^0/\Lambda$ (\broken), $D_q$ (\dotted), $K^+/K^-$ (\full),
$\bar{\Lambda}/\Lambda$ (\chain), $\bar{\Xi}^+/\Xi^-$ (\longbroken)
and the broad $\bar{\Omega}/\Omega$ (\longbroken). The point indicates 
the result of the $\chi^2$-fit in table \ref{sau}.}
\hspace*{2.5cm} \epsfxsize 10cm \epsfbox{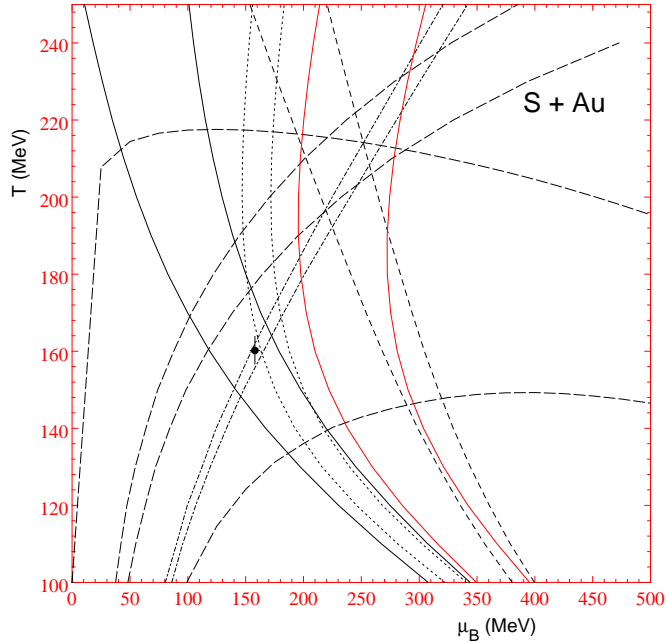}
\end{figure}

We show in figure \ref{saufig} some selected ratios from table 
\ref{sau} together with the point from the $\chi^2$-fit. 
Again we see no perfect agreement but a rather broad region
where various bands concentrate.

The statistical error in the $\chi^2$-fit is determined
by the region where $\chi^2$ increases by one unit from its minimum.
However, from the figures \ref{ssfig}, \ref{sagfig} and \ref{saufig} 
one sees that the systematic error of the model applied
is much larger. From the figures we
conclude that the freeze-out temperature in heavy-ion collisions is 
still very uncertain and one has in fact to take a range of 
$T^{\rm chem} = 150$--200 MeV.
 
The results on sulphur induced collisions at CERN-SPS indicate
no full strangeness equilibration but one is very close to it. 
The slightly higher $\gamma_s$ in the smaller 
collision systems may be a result of having no multi-strange
particles in the $\chi^2$-fit.

\section{Conclusions from thermal models}

Since the application of a statistical model to multi-particle
production by Fermi \cite{Fermi50} its validity is under debate.
Therefore we would first like to make some general remarks.
Thermodynamics is first of all a formalism which may be derived
from statistical (quantum) mechanics in the infinite volume
limit. The basic assumption going in is that all states which
are allowed by conservation laws, including energy conservation,
are equal probable. We get the microcanonical formalism. Going
to the canonical formalism only states sharing the same total energy 
have the same probability which is proportional to 
$\propto \exp(-E/T)$.
The equal probability is usually violated in elementary reactions
where the final state probability is given by the corresponding
matrix element. Therefore the general opinion is that in elementary
reaction the thermal model has no justification. 

However, a large number of particles in the final state suggests
to use a statistical approach. Therefore the thermal model was
applied for elementary reactions
and one gets reasonable agreement \cite{Becattini96a,Becattini97a}.
This observation suggest that if only enough energy is available
the particle production is dominated by the statistical component
and dynamical aspects are of minor importance. In addition Becattini
\cite{Becattini96a,Becattini97a} got the important result that there 
is an universal hadronization 
temperature in different elementary collision systems and 
it is independent on $\sqrt{s}$. The interpretation is
that in the rest frame of the leading particle/parton 
the probability of producing a particle with
energy $E$ is proportional to $\propto \exp(-E/T_h)$. The hadronization
temperature $T_h$ may be identified with the Hagedorn temperature
$T_H$. In the interpretation of Hagedorn \cite{Hagedorn71} it 
is not possible to
create a system of higher temperature than $T_H$ unless there is
a phase transition. The limiting temperature is seen in elementary
reactions even at very high energies like in $p$+$\bar{p}$ at
CERN \cite{Becattini97a}.
The above described observation was also dubbed as 
{\it statistical filling of phase space}. 
We want to point out that there is a difference in the basic
hadron production probability compared to the string models 
for hadronic reactions \cite{Becattini97b}. There the production probability
goes basically like $\propto \exp(-m^2*k)$ \cite{Andersson83}
with $k$ being a universal constant.

The mechanism for chemical equilibration in nucleus-nucleus collisions
is expected to be different from the one in elementary reactions. 
In nuclear reactions 
we assume that chemical equilibration is established by secondary 
interactions among the produced hadrons. Therefore we 
distinguish two mechanisms
which bring the system to maximum entropy or chemical equilibrium.

\begin{itemize}

\item The production of particles, i.e. the hadronization or fragmentation,
      follows a statistical law and already at their production they are
      distributed according to maximum entropy. The ensemble average
      is done by averaging over many events in the experimental analysis.  

\item The maximum of entropy is build up in the classical sense by
      interactions among the particles until detailed balance is reached.
      The ensemble average is reached in each collision by the
      average over the lifetime of the system (engodic theorem). 

\end{itemize}

The thermodynamical formalism cannot distinguish between both 
scenarios and one has to use dynamical arguments to justify one or
the other mechanism. 

Since at SPS-energies the chemical freeze-out temperatures and the 
chemical potentials are very similar in p+p collisions and in 
S+$A$ collisions (see table \ref{summary}) one cannot use them 
to justify secondary interactions for chemical equilibrium.
A superposition of p+p collisions explains most of the
features in the nuclear collisions \cite{Jeon97} with the exception of
strangeness. Therefore we emphasize the importance of the measurement 
of strangeness because there the difference in 
p+p to $A$+$A$ is most clearly seen.

The freeze-out temperature is expected to decrease with increasing $A$, 
because freeze-out occurs when the mean free path is of the order of
the size of the system. Such an effect is not observed in an unambiguous
way, yet. So far only indications are seen as for example the decrease
of chemical freeze-out in our analysis from S+S to S+Ag to S+Au.
A clear sign for chemical equilibration due to
secondary interactions would be a difference in the
chemical freeze-out of p+p compared to a real heavy nucleus like
Pb+Pb or Au+Au. Such an analysis hasn't been performed yet but with the 
now analyzed data of Pb+Pb at CERN-SPS 
(see for example the various contributions to
this proceedings) it will be possible soon. At the AGS we expect
different freeze-out temperatures for Si+Au and Au+Au. The first
results on Au+Au \cite{Cleymans97b} go in this direction but we 
have to wait for the completion of the experimental data analysis.

Coming finally back to strangeness we have here a clear signal of the 
difference between elementary reactions and nuclear collisions.
The strangeness suppression factor $\gamma_s$
is the quantitative measure for the strangeness enhancement
in the thermal model. 
In p+p it is $\gamma_s \approx 0.5$ \cite{Becattini97a} but
in nuclear collisions at CERN-SPS it is around 0.7-1 
(see table \ref{summary}). A strangeness enhancement 
is seen at the AGS, too \cite{Abbott90}, but the situation is not so
clear in terms of the thermal ansatz. A consistent study
of the $\gamma_s$ dependence has not been made. First there is no thermal
analysis of the p+p interaction at the corresponding $\sqrt{s}$
and second the analysis at AGS assumes mostly full
strangeness equilibration as it is seen in table \ref{summary}. 
However, we have already remarked
\cite{Sollfrank95} that a $\gamma_s \approx 0.7$ is better for 
describing the data at AGS.

We have shown that the thermal model fits are not perfect including
all measured particle species. The deviations are a source of
debate. The various interpretations are that the thermal description
is not valid at all, the hadronization of a QGP leaves 
non-equilibrium tracks in special hadron ratios \cite{Rafelski96},
anti-baryons exhibit a large absorption \cite{Spieles95} or
the deviations are not serious \cite{BraunMunzinger96}. The final
answer will be given in the expected high statistic data of Pb+Pb
in the future.

We summarize that there are strong signs of chemical equilibration 
in heavy ion collisions. Since there is already an equally good
chemical equilibrium (excluding strangeness) in the basic p+p collisions 
chemical equilibrium cannot be used for justifying secondary 
hadronic collisions. (There are better signals for abundant secondary 
interactions like the collective flow studies \cite{Bearden97}.)
However, the strangeness production is very different between both 
collision systems and it may be used as the chemometer for 
chemical equilibrium. 

\ack
This work was supported by the Bundesministerium f\"ur Bildung
und Forschung (BMBF) under grand no. 06 BI 556 (6). We gratefully 
acknowledge helpful discussions with H Satz, U Heinz, F Becattini,
M Ga\'zdzicki and J Rafelski. 

\References

\clearpage

\centerline{\large \bf Additional Figure}
\begin{figure} 
\caption{Overview of table \ref{summary} in the $T$-$\mu_B$-plane.
The plot uses color coding and information is lost using
black and white. The plot is not included in the version
appearing in Journal of Physics G}
\end{figure}
\thispagestyle{empty}
\newpage
\hspace*{-1cm}
\epsfxsize 18cm \epsfbox{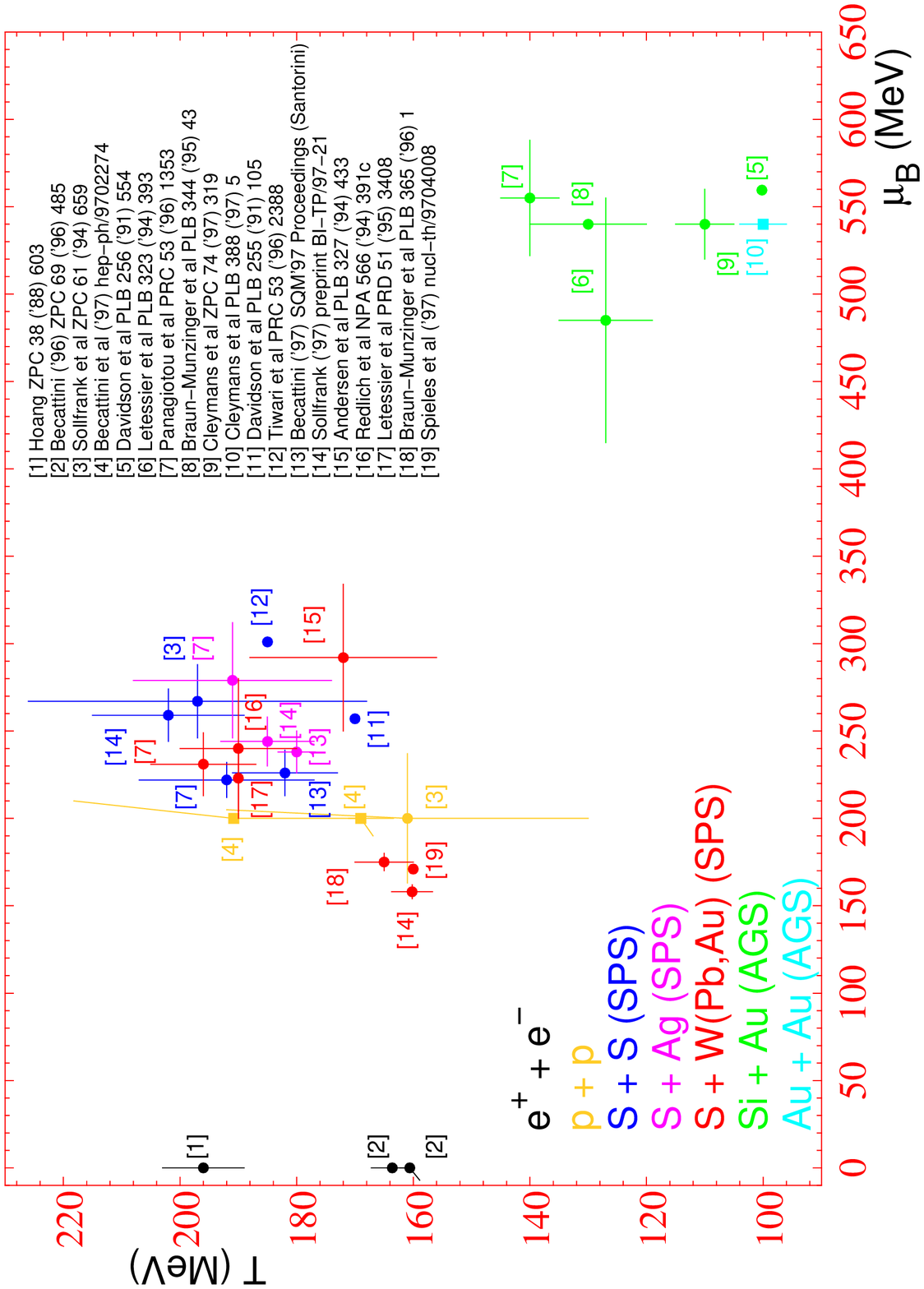}
\thispagestyle{empty}

\end{document}